# Development of A Hardware-In-the-Loop (HIL) Testbed for Cyber-Physical Security in Smart Buildings

**Guowen Li**
*Student Member ASHRAE*

**Zhiyao Yang, PhD**
*Member ASHRAE*

**Yangyang Fu, PhD**
*Member ASHRAE*

**Lingyu Ren, PhD**
*Member ASHRAE*

**Zheng O'Neill, PhD, PE**
*Fellow ASHRAE*

**Chirag Parikh**
*Member ASHRAE*

**ABSTRACT**

*As smart buildings move towards open communication technologies, providing access to the Building Automation System (BAS) through the building's intranet, or even remotely through the Internet, has become a common practice. However, BAS was historically developed as a closed environment and designed with limited cyber-security considerations. Thus, smart buildings are vulnerable to cyber-attacks with the increased accessibility. This study introduces the development and capability of a Hardware-in-the-Loop (HIL) testbed for testing and evaluating the cyber-physical security of typical BASs in smart buildings. The testbed consists of three subsystems: (1) a real-time HIL emulator simulating the behavior of a virtual building as well as the Heating, Ventilation, and Air Conditioning (HVAC) equipment via a dynamic simulation in Modelica; (2) a set of real HVAC controllers monitoring the virtual building operation and providing local control signals to control HVAC equipment in the HIL emulator; and (3) a BAS server along with a web-based service for users to fully access the schedule, setpoints, trends, alarms, and other control functions of the HVAC controllers remotely through the BACnet network. The server generates rule-based setpoints to local HVAC controllers. Based on these three subsystems, the HIL testbed supports attack/fault-free and attack/fault-injection experiments at various levels of the building system. The resulting test data can be used to inform the building community and support the cyber-physical security technology transfer to the building industry.*

## INTRODUCTION

### Background

Serving as the brains for operating smart buildings, modern Building Automation Systems (BASs) include cyber-infrastructure components of sensing, computation, and communication for monitoring and control of the mechanical systems and physical environment in buildings. As the building industry moves towards open communication protocols such as ASHRAE Standard 135-2020 BACnet (ANSI/ASHRAE 2020), providing access to the BAS through the building's intranet, or even remotely through the Internet, has become a common practice. However, the current generation of BAS is designed and operated with limited consideration of the potential cyber vulnerabilities that come with the increased accessibility (Li et al. 2023). As the result, the BAS in smart buildings, especially in the emerging Grid-interactive Efficient Buildings (GEBs), are vulnerable to cyber-attacks that could cause adverse consequences such as occupant discomfort, excessive energy usage, unexpected equipment downtime, and even disruption of grid operation. In order to investigate the BAS behavior under cyber-attacks as well as to develop cyber-resiliency countermeasures for the BAS to safeguard the GEBs, a testbed capable of emulating real-time cyber-physical events in real building operations is highly desired.

The normal operation of GEBs is subject to threats both passively (e.g., physical faults) and actively (e.g., cyber-attacks). These threats can be detected by analyzing the building operation parameters such as zone temperature setpoints as well as network traffic data such as packet length, destination, and transmission rate. The data from normal as well as faulty operations of the BAS under various scenarios are crucial inputs to the development of algorithms for detecting and further mitigating these threats to the GEB. However, such threat data, in particular, network traffic data, is not commonly available. To generate such data for buildings in various operation and threat scenarios, a Hardware-in-the-Loop (HIL) testbed is developed to provide the flexible capability of emulating the complex, expensive, and often less controllable part of the building system with simulations while not compromising the fidelity of the study by incorporating the real physical components in the loop.

The HIL testbed in this study is based on the hardware of the BAS server-controller network to focus on the real network communication and control in modern commercial buildings. Based on an actual BAS server-controller framework, the HIL testbed grants the users full access to the buildings' schedules, setpoints, trends, alarms, and other control functions across the local network. Meanwhile, the building physics along with its HVAC equipment are emulated with simulation since conducting experiments on a real building is a resource-consuming process due to scalability and deployment concerns. The building and HVAC dynamics need to be considered when evaluating and further mitigating the negative effect caused by cyber threats. Besides, the HIL testbed also enables studying the attacks originated from the network and their propagation within the system.

**Contributions**

A HIL testbed is developed with the capability of tracking the real-time network traffic in a typical BAS as well as emulating the building dynamic and HVAC control and operation status. The HIL testbed is based on a standard BACnet compatible BAS in a local network with a server computer and local controllers for HVAC equipment, including chiller, air handling unit (AHU), and variable air volume (VAV) terminal boxes. The BAS server is running on the BACnet/IP network, the controllers are running on the ARCNET network, and a router connects the two networks. The building dynamics, including zone conditions and equipment status, are simulated in real-time via virtual building models on a HIL emulator and directly transmitted to the BAS local controllers. The network traffic between the BAS server computer and the controllers is tracked and available for analysis. The normal and fault operation status data and network traffic data generated by the HIL testbed can be used to develop algorithms for detecting and mitigating cyber-attacks. Such a HIL testbed can also be utilized for other sensor and control-related studies, including demand flexibility (Li et al. 2022; Chen et al. 2019), occupant-centric controls, smart and connected communities, etc.

## HIL TESTBED

### Configuration

Figure 1 presents the HIL testbed at Texas A&M University (TAMU) campus. The HIL testbed consists of three major components: a BAS server computer, a real-time HIL emulator, and multiple local BAS equipment controllers.

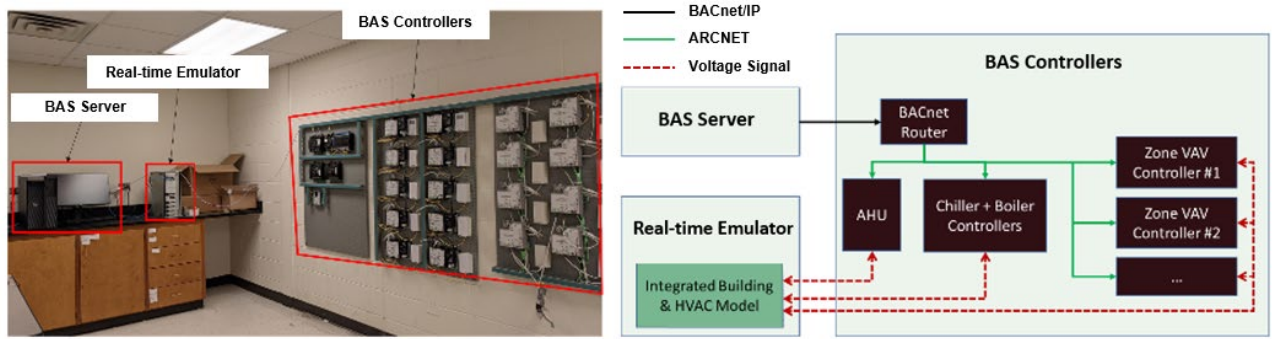

**Figure 1** The HIL testbed developed at TAMU.

The BAS server computer oversees all BAS controllers' operation status and provides real-time supervisory control capability. The BAS server is connected to the BAS local controllers via Ethernet following the BACnet/IP protocol. The BAS server can be used to make changes to the default control logic of each local equipment controller. The server as a workstation also hosts the software environment for the testbed, including an open-source database that supports data storing and querying, a set of the HIL machine software tools such as ControlDesk that controls the HIL experiment, and an adversary program that performs cyber-attacks on the building control system utilizing the vulnerabilities of BACnet protocol.

The BAS controllers contain the control logic of the associated equipment from Standard Application Library (SAL) file, and in real buildings, they are normally installed close to the equipment under control. The HIL testbed includes several types of HVAC equipment (e.g., chiller, AHU, VAV box) controllers. These controllers are all wired to the BACnet/IP to ARCNET router in the daisy-chain connection. During normal operation, these controllers take in signals from local sensors and output control signals accordingly. They also respond to the supervisory control dispatched directly by the BAS server, and the control outputs of the local controllers are accessible in the BAS server. Both the measurement inputs and the control outputs of a local controller can be connected directly to outside sources with voltage input/output (I/O). The default control logic in these controllers can be updated in the BAS server and downloaded to the local control board.

The HIL machine (i.e., real-time emulator) provides real-time emulation of a building energy system modeled in Modelica. The building model takes the control commands from BAS controllers as inputs and generates typical measurements for the building energy system as outputs that are sent back to the BAS controllers. The real-time emulation depends on a set of hardware and software. For the hardware, additional digital/analog (D/A), A/D boards, and wires are used to establish the communications between the emulator and the BAS controllers. For example, the simulation models are compiled and deployed to the emulator with designated channels receiving signals such as a damper position from the BAS controllers via the A/D boards and used as the input to the simulation models. The simulated results, such as the zone temperature, are then transmitted back to the BAS controllers via the D/A boards. With the real-time emulator, the operation of the building and its HVAC system can be directly emulated without the need for actual zones and equipment.

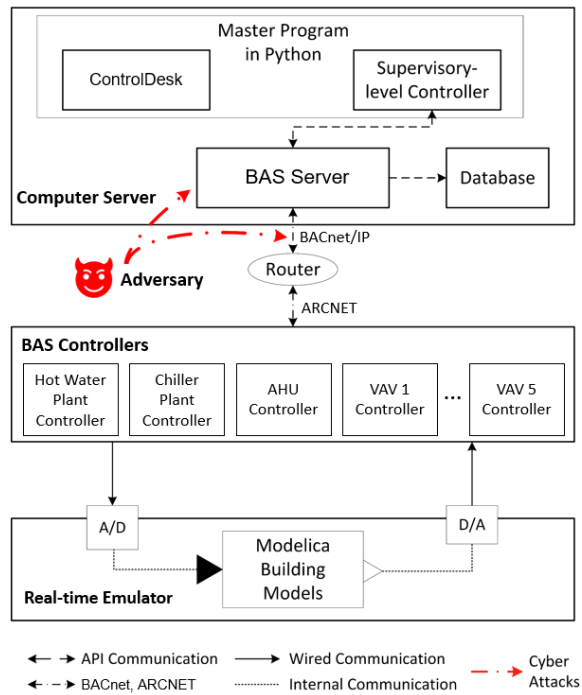

**Figure 2** Data transmission schematic in the HIL testbed.

## Dataflow

Figure 2 describes the detailed dataflow for the HIL testing. The computer server hosts the software environment for all the hardware and customized services, such as a master program that controls the experiments. A BAS server is connected to the local controllers through the BACnet/IP to ARCNET router. The daisy-chain connection is established between the router and each local controller. The BAS controllers provide rule-based control logic for different HVAC equipment, whose control commands are sent via voltage signals to the virtual building in the real-time emulator. The real-time emulator emulates a virtual building that is modeled in Modelica (Fritzson 2014).

# EXPERIMENTAL DESIGN

## Virtual Building Model

Figure 3 illustrates the schematic diagram of a typical HVAC system for a medium-sized office building. Heating and cooling are delivered by a single-duct VAV system. One AHU connected with five VAV terminal boxes serves five zones (four perimeter zones and one interior zone) on one floor. The chilled water is supplied by a central chiller plant that consists of a chiller, a waterside economizer, a cooling tower, a chilled water pump, and a cooling water pump. A boiler, fed by natural gas, supplies the hot water to the AHU heating coils.

The virtual building model is modeled based on the open-source Modelica Buildings Library (MBL) developed by Lawrence Berkeley National Laboratory (Wetter et al. 2014). Modelica, an equation-based modeling language, supports accurately and flexibly modeling and simulating building energy and control systems for control-oriented studies (Li et al. 2021). This Modelica model is based on and validated against the medium-size office prototype model developed by Pacific Northwest National Laboratory in EnergyPlus (Goel et al. 2014). As shown in Figure 4, the system model consists of an HVAC system, a building envelope model, and a model for air flow through building leakage and through open doors based on wind pressure and flow imbalance of the HVAC system. The HVAC system is sized for Chicago, IL in climate zone 5A. The air-side control sequences follow ASHRAE Guideline 36 (ASHRAE 2018), and the water-side control sequences follow ASHRAE project RP-1711 (Taylor 2020). More details of this HVAC system can be found in (Fu et al. 2021).

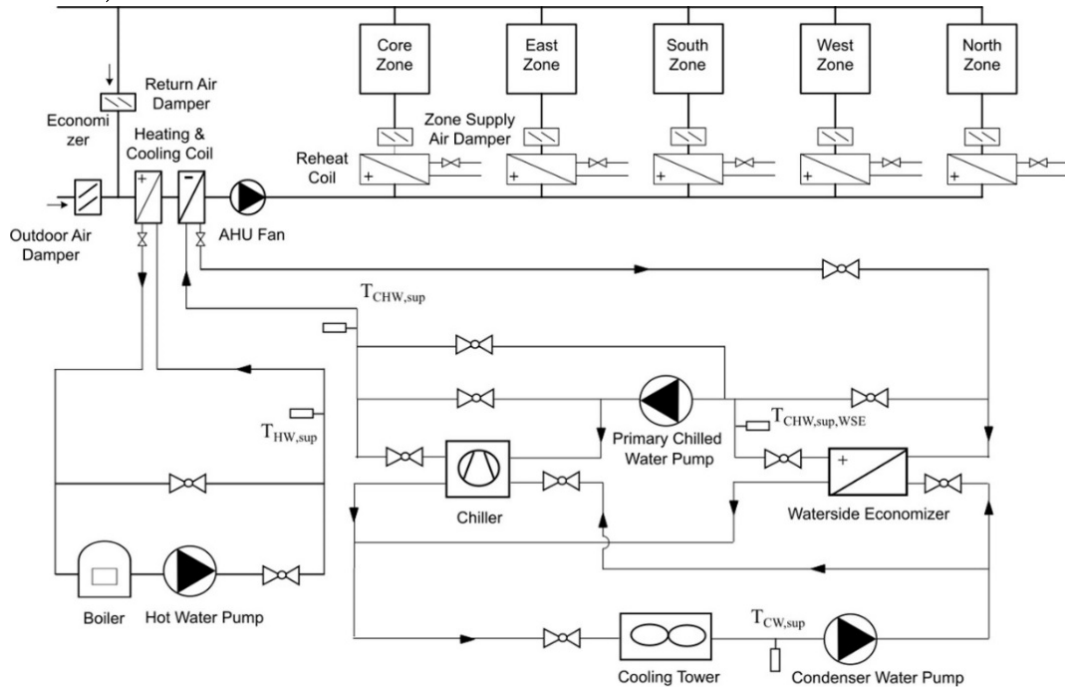

**Figure 3** The studied HVAC system in the real-time HIL emulator (Fu, O'Neill, and Adetola 2021).

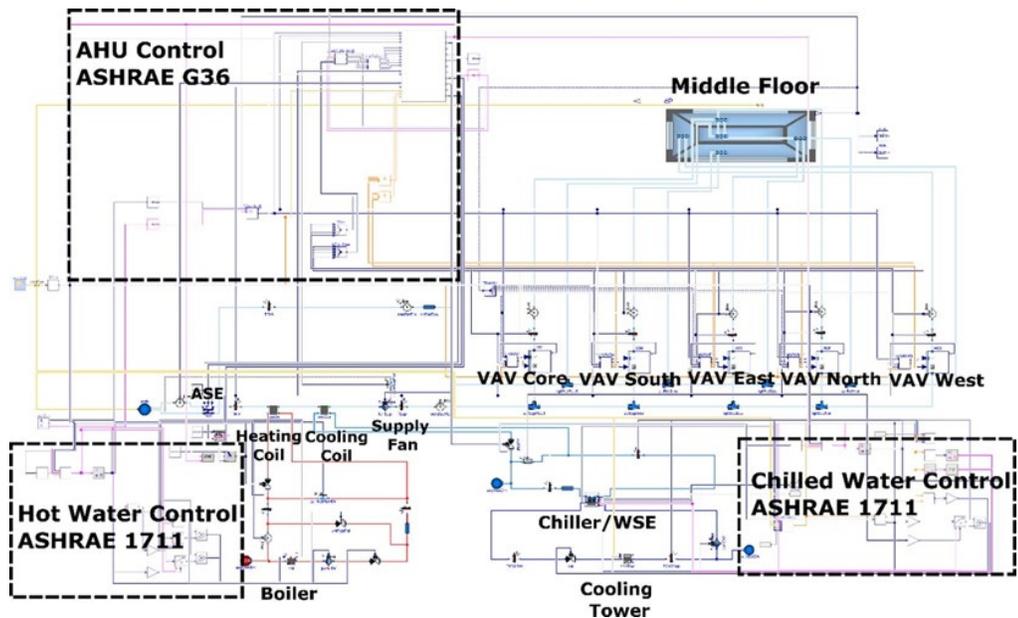

**Figure 4** Modelica implementation of the studied HVAC system (Fu et al. 2021).

**False Data Injection Attack Dataset**

In the following experiment, we assume the attacker gained credentials to the BAS server through a brute-force attack. The attacker then leveraged the Python interface to remotely send malicious setpoints to the BAS server. To create the most severe impact, we changed the temperature setpoint of AHU supply air to its maximum value, 95 °F, from 10:00 am to 11:00 am. This scenario was tested on August 1st using Typical Meteorological Year, version 3 (TMY3) weather datasets of Chicago. Figure 5 shows the experimental data of the false data injection attack scenario. The black lines indicate the normal operating data, which represents the baseline that was free of faults and attacks. For the cooling days, the zone temperature is controlled at 75°F during occupied periods by adjusting the zone supply air flowrate and AHU supply air temperature in response to zone heat gains and outdoor air conditions. The occupied schedule is 7:00 am – 8:00 pm. The red lines indicate the faulty data measured from the HIL testbed. It's noted that the fault injection impacts the building system during both the attack period and the post-attack period. The post-attack period is defined as the time that a system requires to recover to its baseline operation after the attack (Fu, O'Neill, and Adetola 2021). For example, during the attack period (10:00 am to 11:00 am), both AHU supply air temperature and zone temperature rose up to 80 °F (26.67 °C). Then AHU supply temperature and zone temperature returned to the baseline during the post-attack period (11:00 am to 2:00 pm).

**Device Denial-of-Service Attack Dataset**

In most applications, the BACnet protocol does not require authentication for field devices, nor does it encrypt the payloads. An attacker device could register on the BACnet/IP router as a foreign device and join the local broadcast list. To interrupt the service of a critical field device, the attacker can keep sending reinitialization requests so that the target device is constantly in soft-rebooting and fails to answer any benign requests. We applied this device denial-of-service (DoS) attack on the AHU controller from 10:00 am to 11:30 am. This scenario was also tested on August 1st using TMY3 weather datasets of Chicago. Figure 6 shows the experimental data of the device DoS attack scenario. The red lines indicate the faulty data measured from the HIL testbed. The black lines represent the baseline that was free of faults and attacks. During the experiment, it was observed that the device DoS attack caused missing data points to certain variables of the AHU controller during the attack period. It's noted that the cyber-attack impacted the building system during the attack period (10:00 am to 11:30 am) as well as the post-attack period (11:30 am to 3:00 pm).

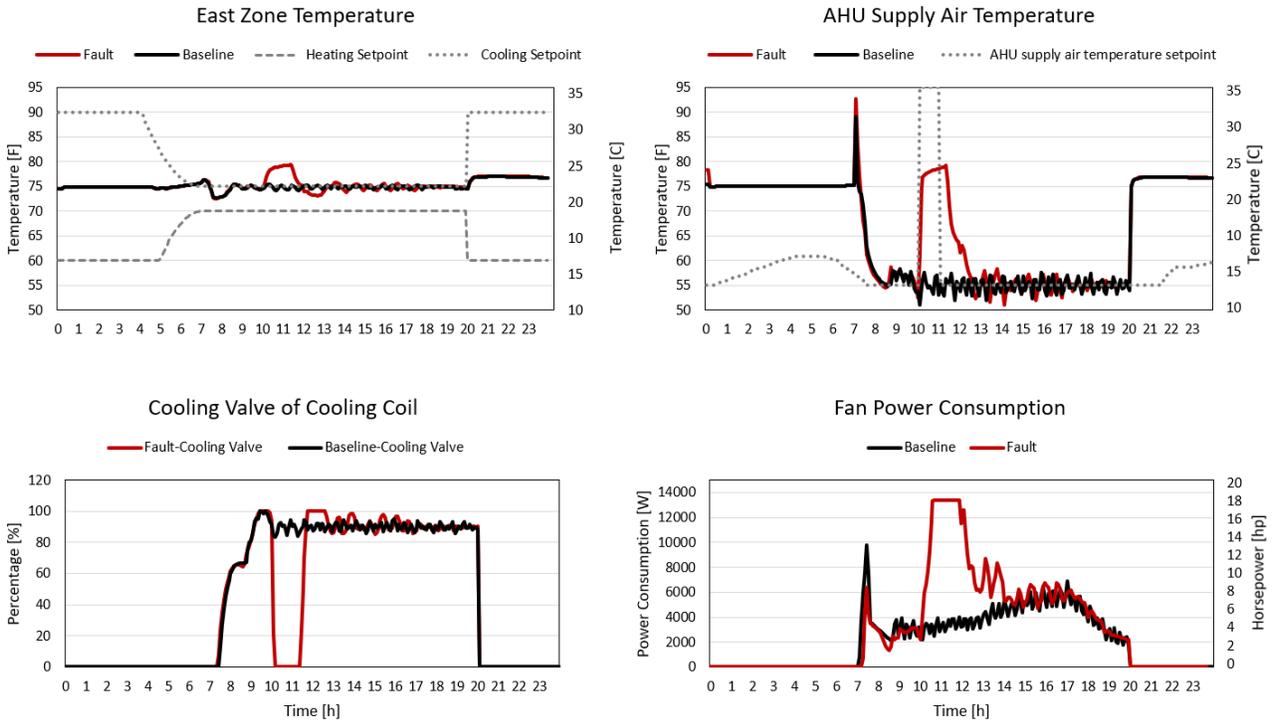

**Figure 5**　Measurement data for the false data injection attack scenario.

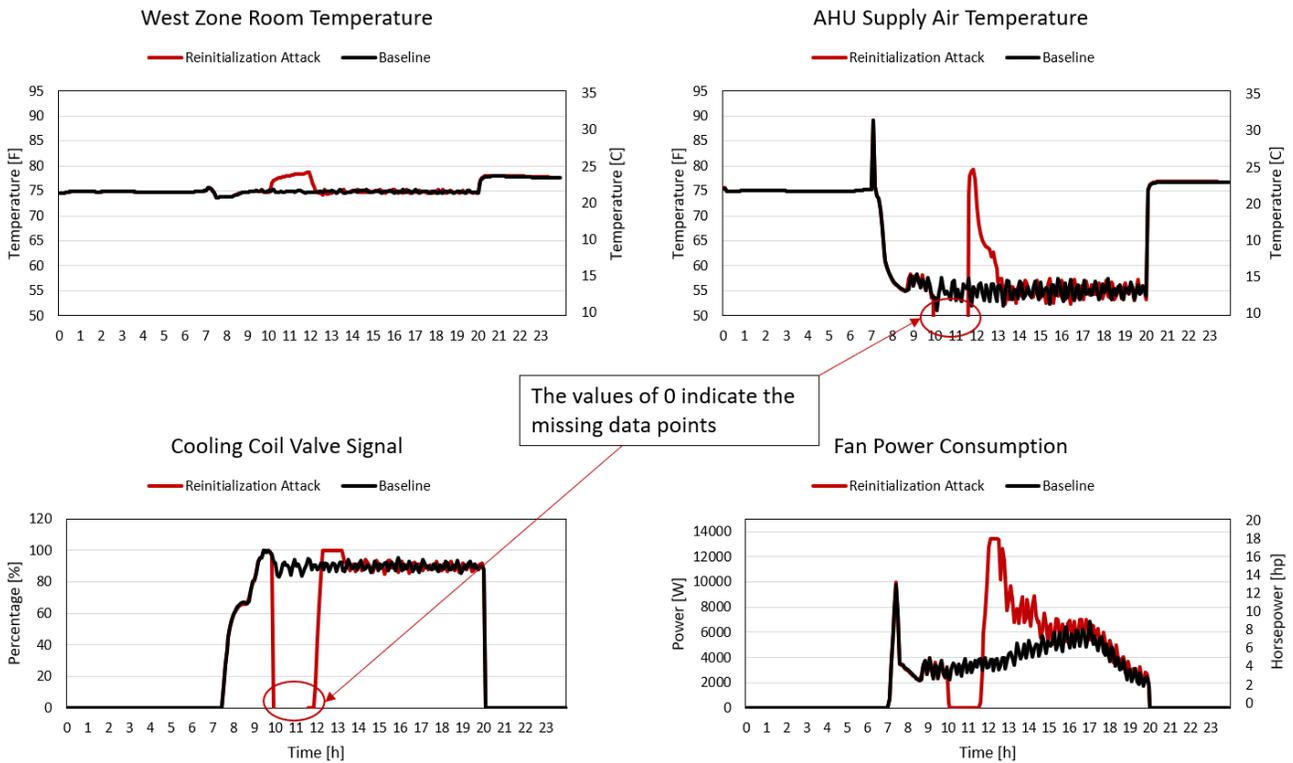

**Figure 6**　Measurement data for the device DoS attack scenario.

**Network Traffic Monitoring**

    Besides the BAS operating data recorded by the BAS server, the network traffic in the BAS local network is also critical data used to detect cyber threats. Therefore, the HIL testbed is also equipped with network traffic monitoring capabilities to generate the data with sufficient details for threat analytics. Using the open-source Wireshark software on the BAS server computer enables network traffic monitoring on the Ethernet port connecting the server and the BACnet router and the rest of the BAS local controllers. Figure 7 and Figure 8 show screenshots of the network traffic details and packets. The Wireshark keeps track of the addresses, length, protocol, and related information of each packet. Packets are transmitted between the BAS server (10.13.254.2) and the BACnet router (10.13.254.5) using the BACnet/IP protocol. These detailed network traffic data generated by the HIL testbed are instrumental for developing algorithms to detect irregular network activities and identify cyber threats against the BAS.

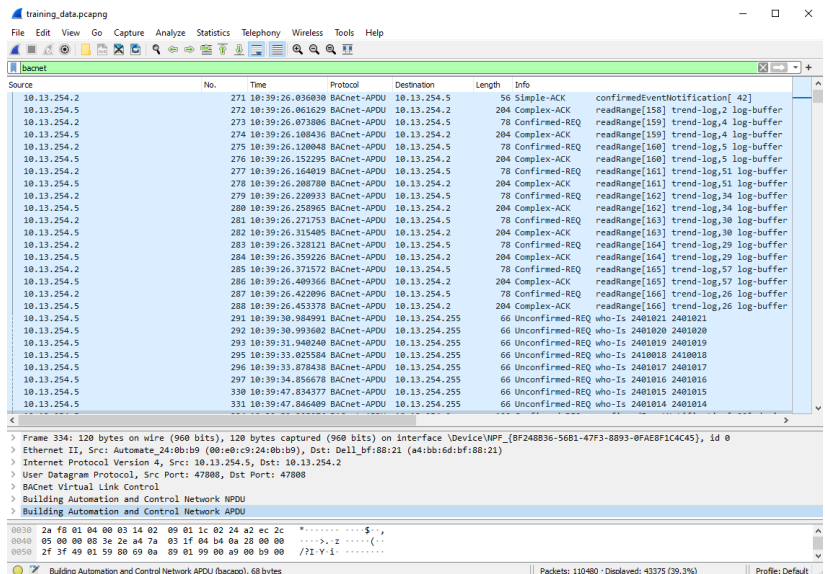

**Figure 7**   Network traffic details recorded by Wireshark.

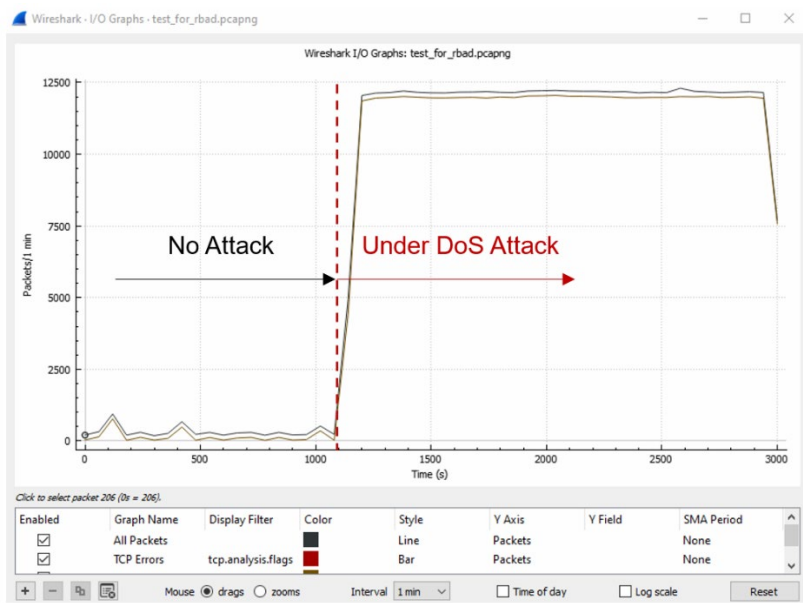

**Figure 8**   Network packets visualized by Wireshark.

## CONCLUSIONS AND FUTURE WORK

This paper presented a flexible HIL testbed developed at TAMU for cyber-physical security in smart buildings. The HIL testbed consists of a real-time emulator, a set of real BAS controllers, and a BAS server computer. The hardware setup and data transmission within the HIL testbed are described in detail. The data generation capability of the HIL testbed is demonstrated by tracking the normal and faulty operating data in the BAS, as well as monitoring the detailed network traffic in the local BAS network. The fully functional HIL testbed is being used to facilitate the development of threat detection and mitigation algorithms for cyber-secured buildings. Such a HIL testbed can also be utilized for other sensor and control-related studies, including demand flexibility, occupant-centric controls, smart and connected communities, etc.

## ACKNOWLEDGMENTS AND DISCLAIMER

The research reported in this paper was partially supported by the Building Technologies Office at the U.S. Department of Energy through the Emerging Technologies program under award number DE-EE0009150.